\documentclass[twocolumn,nofootinbib,preprintnumbers,amsmath,amssymb]{revtex4-2}

\usepackage{graphicx}
\usepackage{dcolumn}
\usepackage{bm}
\usepackage[usenames]{color}
\usepackage{graphicx,epsfig}
\usepackage[colorlinks={true}]{hyperref}
\hypersetup{colorlinks=true,linkcolor=red,citecolor=blue,urlcolor=blue}
\usepackage{orcidlink}
\usepackage{braket}

\begin{document}

\title{Quantum Capacitor: A Coherence-Based Quantum Energy Storage Device}

\author{Saeed~Haddadi\,\,\orcidlink{0000-0002-1596-0763}}
\email{haddadi@ipm.ir}
\affiliation{
School of Particles and Accelerators,\\
Institute for Research in Fundamental Sciences (IPM),\\
P.O. Box 19395-5531, Tehran, Iran
}


\begin{abstract}
Quantum batteries have recently emerged as promising candidates for microscopic energy-storage technologies exploiting uniquely quantum mechanical effects. In this work, we introduce the concept of a \emph{quantum capacitor}, a coherence-based quantum energy-storage device exhibiting reversible and ultrafast charging-discharging dynamics. Unlike conventional quantum batteries, which primarily focus on extractable work, the proposed quantum capacitor operates through reactive energy accumulation mediated by coherent quantum polarization. We formulate a theoretical framework based on a driven two-level system and define a dynamical quantum capacitance associated with the response of stored energy to coherent external driving. The charging power, reversible energy exchange, and the effects of environmental decoherence are investigated analytically within the Lindblad formalism. We extend the proposal to interacting many-body systems and show that collective coherence and interactions can enhance the effective quantum capacitance and charging power, potentially leading to cooperative energy-storage advantages beyond the single-qubit regime.
\end{abstract}

\maketitle

\section{Introduction}
The rapid development of quantum technologies has generated considerable interest in the possibility of storing, transferring, and manipulating energy at the quantum scale. In recent years, the concept of a \emph{quantum battery} has emerged as a promising framework for microscopic energy-storage devices capable of exploiting uniquely quantum mechanical resources such as coherence, entanglement, and collective many-body interactions~\cite{campaioli2024colloquium}. Unlike classical batteries, quantum batteries may exhibit charging advantages originating from nonclassical correlations and cooperative quantum effects, potentially enabling ultrafast charging processes and enhanced energy-transfer efficiencies~\cite{alicki2013entanglement,PhysRevLett.118.150601,binder2015quantacell,ferraro2018high,PhysRevLett.122.047702}.
The study of quantum batteries has rapidly evolved into a central topic in quantum thermodynamics and nonequilibrium quantum physics. Numerous investigations have explored collective charging protocols, quantum speed limits, ergotropy extraction, many-body charging advantages, and the role of entanglement in enhancing charging power~\cite{PhysRevLett.125.040601,quach2023quantum,quach2020organic,kamin2020entanglement,shi2022entanglement,ahmadi2024nonreciprocal,rossini2019many,yang2023battery}. Moreover, some progress in superconducting circuits~\cite{hu2022optimal,PhysRevA.107.023725,AsadAli2025}, cavity-based systems~\cite{HaddadiQB2024}, trapped ions~\cite{prlti}, magnetic dipolar systems~\cite{Ali2025Asad}, and spin-chain architectures~\cite{dou2022cavity,yao2022optimal,zhang2023quantum,ali2024ergotropy} have significantly increased the feasibility of realizing microscopic quantum energy-storage devices.

Despite these advances, most existing studies focus almost exclusively on battery-like quantum systems whose primary purpose is maximizing stored work or extractable energy. In contrast, classical energy-storage technologies are fundamentally divided into two distinct categories: batteries and capacitors. While batteries typically rely on relatively slow chemical storage processes optimized for long-term energy retention, capacitors operate through reversible electrostatic energy accumulation and are characterized by ultrafast charging and discharging cycles. Capacitors are therefore essential in modern electronics whenever rapid energy delivery, transient power stabilization, and low-dissipation operation are required.
Thus, the notion of a genuinely \emph{quantum capacitor} (QC) remains comparatively unexplored within the context of quantum thermodynamics and quantum energy-storage theory. Existing literature on quantum capacitance~\cite{Luryi1988,PhysRevB.100.075433,kitsenko2026} primarily concerns mesoscopic transport phenomena, graphene systems, double quantum dots, superconducting circuits, or quantum RC devices, where the term ``quantum capacitance'' refers to density-of-states effects and electronic compressibility. These approaches differ substantially from the idea of a coherence-assisted quantum energy-storage device operating analogously to a classical capacitor.
Motivated by this distinction, in the present work we introduce the concept of a QC: a quantum energy-storage device whose operation is governed by reversible coherent dynamics and rapid energy exchange processes. In contrast to conventional quantum batteries, the proposed QC emphasizes reactive energy storage mediated by coherent quantum polarization. Energy is accumulated temporarily through coherent oscillatory dynamics and may subsequently be released on ultrafast timescales with minimal entropy production.

The physical mechanism underlying the QC is fundamentally connected to quantum coherence~\cite{baumgratz2014quantifying}. Coherent superposition between energy eigenstates enables oscillatory energy accumulation analogous to charge accumulation in classical capacitors. Consequently, the QC behaves as a reactive quantum element rather than a dissipative storage device. Such behavior naturally establishes a bridge between quantum thermodynamics, quantum coherence theory, and quantum transport phenomena.
Another important distinction between quantum batteries and the proposed QC lies in the operational timescale. Quantum batteries are typically designed to maximize total extractable work, whereas the QC is specifically optimized for rapid charging-discharging cycles and reversible energy exchange. This feature may become particularly relevant for nanoscale quantum technologies requiring transient power delivery, ultrafast switching operations, or low-dissipation energy redistribution.

In this work, we formulate a theoretical description of the QC based on a coherently driven two-level system. We introduce a definition of quantum capacitance associated with the susceptibility of stored energy to external driving fields and investigate charging-discharging dynamics governed by coherent oscillations. Furthermore, the effects of environmental decoherence are analyzed within the Lindblad formalism. Our results demonstrate that coherence-induced oscillatory dynamics naturally provide capacitor-like energy-storage behavior at the quantum scale. We extend the current framework to many-body QCs, where collective coherence and entanglement can significantly increase the effective quantum capacitance and lead to extremely large scaling of reversible energy storage and charge rates beyond the single-qubit regime.

The paper is organized as follows. In Sec.~\ref{SECII}, we introduce the driven two-level model describing the QC. Section~\ref{SECIII} presents the definition of quantum capacitance and discusses its physical interpretation. In Sec.~\ref{SECIV}, we analyze charging and discharging dynamics, while Sec.~\ref{SECV} investigates decoherence effects. Many-body QC extension and possible physical implementations are discussed in Secs.~\ref{SECVI} and \ref{SECVII}, followed by concluding remarks in Sec.~\ref{SECVIII}.

\section{Model}
\label{SECII}
\begin{figure}
  \centering
  \includegraphics[width=80mm]{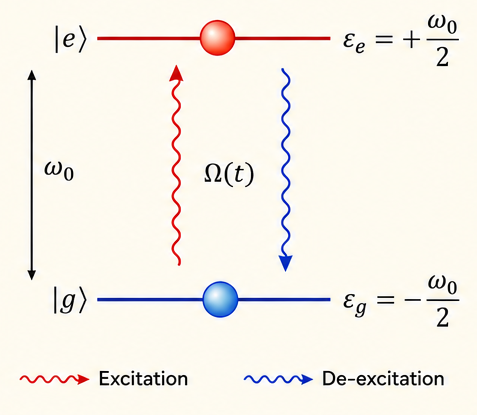}
  \caption{Schematic of a driven two-level system interacting with an external charging field.}\label{Fig:1}
\end{figure}

To capture the essential physics of a QC, we consider the simplest nontrivial quantum system capable of storing energy coherently, namely a driven two-level system interacting with an external charging field (Fig.~\ref{Fig:1}). The two-level configuration provides a minimal platform for investigating reversible energy accumulation through coherent quantum dynamics and allows direct analytical treatment of the charging process.
The total Hamiltonian of the QC can be written as
\begin{equation}
H = H_0 + H_{\mathrm{int}},
\label{totalham}
\end{equation}
where
$H_0=\frac{\omega_0}{2}\sigma_z$
describes the internal energy of the isolated quantum system and
$H_{\mathrm{int}}=\Omega(t)\sigma_x$
represents the interaction with the external coherent charging field.
Here, $\omega_0$ denotes the intrinsic transition frequency between the ground state $\ket{g}$ and excited state $\ket{e}$, while $\Omega(t)$ is the time-dependent driving amplitude controlling the charging process. The Pauli operators are defined as
$\sigma_z=\ket{e}\bra{e}-\ket{g}\bra{g}$
and
$\sigma_x=\ket{e}\bra{g}+\ket{g}\bra{e}$.
The free Hamiltonian $H_0$ possesses two eigenvalues,
$\varepsilon_e=+\omega_0/2$
and
$\varepsilon_g=-\omega_0/2$,
corresponding to the excited and ground states, respectively.
Initially, the system is assumed to be prepared in its ground state,
$\rho(0)=\ket{g}\bra{g}$,
such that the QC initially contains no stored energy. Once the coherent driving field is activated, transitions between $\ket{g}$ and $\ket{e}$ become possible, leading to coherent superpositions of the form
$\ket{\psi(t)}
=
c_g(t)\ket{g}
+
c_e(t)\ket{e}$,
where $c_g(t)$ and $c_e(t)$ are the corresponding probability amplitudes satisfying
$|c_g(t)|^2+|c_e(t)|^2=1$.
The emergence of coherent superposition states plays a central role in the operation of the QC. Unlike classical batteries, where energy is stored through irreversible chemical processes, the QC accumulates energy through reversible coherent oscillations between quantum states. So, the stored energy remains dynamically exchangeable and may be released rapidly through controlled coherent evolution.

The time evolution of the system is governed by the Schr\"odinger equation,
$i\frac{d}{dt}\ket{\psi(t)}
=
H\ket{\psi(t)}$.
For constant driving amplitude $\Omega(t)=\Omega$, the Hamiltonian becomes time independent and can be diagonalized analytically. The corresponding generalized Rabi frequency is given by
$\Omega_R
=
\sqrt{
\Omega^2+\frac{\omega_0^2}{4}
}$.
Under coherent evolution, the excited-state occupation probability oscillates periodically according to
\begin{equation}
P_e(t)
=
\frac{\Omega^2}{\Omega_R^2}
\sin^2(\Omega_R t).
\label{occupation}
\end{equation}
These oscillations constitute the microscopic origin of the capacitor-like behavior of the QC. In direct analogy with periodic charge accumulation in classical capacitors, the coherent oscillatory population transfer between quantum levels generates reversible energy-storage cycles.

The instantaneous stored energy is defined as the expectation value of the internal Hamiltonian relative to the initial ground-state energy,

\begin{equation}
E(t)
=
\textmd{Tr}[\rho(t)H_0]-E(0),
\label{storedenergy}
\end{equation}
where
$\rho(t)=\ket{\psi(t)}\bra{\psi(t)}$
is the density matrix of the system. Using Eq.~(\ref{occupation}), the stored energy becomes

\begin{equation}
E(t)
=
\omega_0
\frac{\Omega^2}{\Omega_R^2}
\sin^2(\Omega_R t).
\label{energyfinal}
\end{equation}
Above equation clearly demonstrates the oscillatory charging-discharging dynamics characteristic of the QC. The energy periodically increases and decreases due to coherent quantum evolution, reflecting the reactive nature of the storage mechanism.
An important feature of the proposed model is the direct role of quantum coherence in the charging process. The off-diagonal density matrix elements,
$\rho_{eg}(t)
=
c_e(t)c_g^*(t)$,
quantify the coherence generated during charging. These coherence terms are responsible for reversible energy exchange and therefore play a role analogous to polarization in classical capacitors.

\section{Quantum Capacitance}
\label{SECIII}
A central quantity in the proposed framework is the notion of \emph{quantum capacitance}, which characterizes the ability of a quantum system to store energy under coherent external driving. In classical electrodynamics, capacitance quantifies the proportionality between stored charge and applied voltage, or equivalently, the susceptibility of stored electrostatic energy to an external potential difference. For a conventional capacitor, the stored energy is given by

\begin{equation}
E=\frac{1}{2}CV^2,
\label{classicalenergy}
\end{equation}
where $C$ is the capacitance and $V$ denotes the applied voltage.

In the present quantum framework, the external coherent driving field plays a role analogous to the voltage source in classical capacitors. The charging field induces coherent transitions between quantum energy levels, thereby generating reversible energy accumulation. Hence, the storage capability of the system should be associated with the response of the stored energy to variations in the external driving strength.
Motivated by this analogy, we define the \emph{quantum capacitance} of the QC as

\begin{equation}
C_Q
=
\frac{\partial E}{\partial \Omega},
\label{cqdef}
\end{equation}
where $E$ denotes the stored energy and $\Omega$ is the coherent driving amplitude. This definition measures the sensitivity of the stored energy to the external charging field and therefore quantifies the energy-storage susceptibility of the quantum system.

Using the analytical expression for the stored energy obtained in Eq.~(\ref{energyfinal}),
the quantum capacitance becomes
\begin{align}
C_Q
=
&
\,
\omega_0
\frac{2\Omega(\Omega_R^2-\Omega^2)}
{\Omega_R^4}
\sin^2(\Omega_R t)
+
\omega_0
\frac{\Omega^3 t}
{\Omega_R^3}
\sin(2\Omega_R t).
\label{cqexplicit}
\end{align}
This expression~(\ref{cqexplicit}) reveals several important physical properties of the QC.
First, the quantum capacitance is intrinsically time dependent due to coherent oscillatory dynamics. Unlike classical capacitors, whose capacitance is typically fixed by geometry and dielectric properties, the QC exhibits dynamical capacitance governed by quantum evolution.
Second, the quantum capacitance depends explicitly on the coherent driving amplitude. This demonstrates that the storage capability of the QC is not solely determined by intrinsic system parameters but can also be controlled externally through coherent driving protocols.
Third, the oscillatory behavior of $C_Q$ reflects the reversible nature of energy exchange between the QC and the external field. Positive values of $C_Q$ correspond to charging processes in which energy is accumulated, while decreasing or oscillatory behavior indicates reversible energy release.

To obtain additional physical insight, it is useful to analyze the weak-driving regime,
$\Omega \ll \omega_0$.
In this limit, the generalized Rabi frequency simplifies to
$\Omega_R
\approx
\omega_0/2$,
and the stored energy becomes
\begin{equation}
E(t)
\approx
\frac{4\Omega^2}{\omega_0}
\sin^2\left(
\frac{\omega_0 t}{2}
\right).
\label{weakenergy}
\end{equation}
So, the quantum capacitance reduces to
\begin{equation}
C_Q
\approx
\frac{8\Omega}{\omega_0}
\sin^2\left(
\frac{\omega_0 t}{2}
\right).
\label{weakcq}
\end{equation}
Equation~(\ref{weakcq}) shows that the quantum capacitance scales linearly with the driving amplitude in the perturbative regime. Therefore, stronger coherent driving fields enhance the energy-storage response of the QC.

An important distinction between the proposed QC and conventional quantum batteries emerges naturally from the behavior of $C_Q$. Standard quantum batteries are usually characterized by quantities such as ergotropy, charging power, or extractable work~\cite{campaioli2024colloquium,allahverdyan2004maximal}. In contrast, the QC is characterized by its reversible storage susceptibility and oscillatory energy response. The quantum capacitance therefore quantifies the reactive component of the energy-storage dynamics rather than the total extractable work alone.
Table~\ref{table1} summarizes the fundamental differences between conventional quantum batteries and the proposed QC.


\begin{table}[t]
\caption{Quantum battery vs quantum capacitor.}
\label{table1}
\begin{ruledtabular}
\begin{tabular}{p{2.3cm} p{3.0cm} p{3.0cm}}
Property & Battery & Capacitor \\ \hline

Objective &
Work (ergotropy) &
Reactive energy \\

Mechanism &
Population inversion &
Coherent polarization \\

Resource &
Entanglement, coherence &
Coherence, superposition \\

Exchange &
Unidirectional &
Periodic reversible \\

Figure &
Ergotropy; power &
Capacitance; power \\

Timescale &
Max work &
Ultrafast cycles \\

Coherence role &
Enhancement &
Storage basis \\

Many-body &
Collective charging &
Collective capacitance \\

Application &
Work storage &
Fast redistribution
\end{tabular}
\end{ruledtabular}
\end{table}


\section{Charging and Discharging Dynamics}
\label{SECIV}
The most distinctive feature of the proposed QC is its reversible charging-discharging behavior governed by coherent quantum evolution.
As discussed in the previous section, the quantum capacitance $C_Q$ characterizes the susceptibility of the stored energy to coherent driving. So, the charging and discharging dynamics of the QC are directly controlled by the coherent oscillatory behavior responsible for the emergence of $C_Q$. In particular, the periodic variation of the stored energy naturally induces time-dependent charging and discharging cycles analogous to those observed in classical capacitive systems.

From Eq.~(\ref{energyfinal}), we find that the energy stored in the QC evolves periodically in time due to coherent population transfer between the ground and excited states.
The charging stage corresponds to intervals satisfying

\begin{equation}
\frac{dE(t)}{dt}>0,
\end{equation}
during which energy flows from the external driving field into the quantum system. Conversely, the discharging regime is characterized by

\begin{equation}
\frac{dE(t)}{dt}<0,
\end{equation}
indicating that the stored energy is released back from the QC.
Differentiating Eq.~(\ref{energyfinal}), the instantaneous power becomes

\begin{equation}
P(t)
=
\frac{dE(t)}{dt}
=
\omega_0
\frac{\Omega^2}{\Omega_R}
\sin(2\Omega_R t).
\label{power}
\end{equation}
The oscillatory sign of $P(t)$ clearly demonstrates the reversible nature of the energy exchange process. Positive values of $P(t)$ correspond to charging processes, whereas negative values indicate discharging dynamics.

The maximum charging power is obtained when
$\sin(2\Omega_R t)=1$,
yielding

\begin{equation}
P_{\max}
=
\omega_0
\frac{\Omega^2}{\Omega_R}.
\label{maxpower}
\end{equation}
Equation~(\ref{maxpower}) reveals that the charging power increases with the driving amplitude $\Omega$, indicating that stronger coherent fields accelerate the energy-transfer process.

An important feature of the QC is the existence of fully reversible charging cycles. Maximum stored energy is reached at times satisfying

\begin{equation}
\Omega_R t
=
\frac{\pi}{2},
\frac{3\pi}{2},
\dots
\end{equation}
for which

\begin{equation}
E_{\max}
=
\omega_0
\frac{\Omega^2}{\Omega_R^2}.
\label{maxenergy}
\end{equation}

Another important characteristic of the QC is the charging timescale. The characteristic charging time can be estimated from the first energy maximum,

\begin{equation}
\tau_c
=
\frac{\pi}{2\Omega_R},
\label{timescale}
\end{equation}
which demonstrates that increasing the coherent driving amplitude decreases the charging time and enables ultrafast energy accumulation. This feature may become particularly advantageous for quantum technologies requiring rapid transient power delivery or fast energy redistribution.

In the weak-driving regime $(\Omega \ll \omega_0)$, the charging time approximately reduces to

\begin{equation}
\tau_c
\approx
\frac{\pi}{\omega_0},
\end{equation}
while in the strong-driving regime $(\Omega \gg \omega_0)$,

\begin{equation}
\tau_c
\approx
\frac{\pi}{2\Omega}.
\end{equation}
Therefore, strong coherent driving substantially enhances the charging speed of the QC.

The reversible oscillatory dynamics also establish a direct connection between charging behavior and quantum capacitance. Since $C_Q$ measures the sensitivity of stored energy to coherent driving, large values of $C_Q$ correspond naturally to rapid charging rates and enhanced energy responsiveness. The charging and discharging processes may therefore be viewed as dynamical manifestations of the underlying quantum capacitance discussed in the previous section.
From a broader perspective, the QC behaves as a reactive quantum element whose operation resembles the behavior of classical LC or RC circuits, although the underlying mechanism is entirely quantum mechanical. The coherent oscillatory exchange of energy may eventually provide the basis for microscopic quantum energy networks and coherent quantum electronic architectures.

\section{Effect of Decoherence}
\label{SECV}
The operation of the proposed QC relies fundamentally on coherent quantum evolution. As discussed in the previous sections, the reversible charging-discharging dynamics and the emergence of quantum capacitance originate from coherent superpositions between quantum energy eigenstates. Consequently, environmental interactions and decoherence processes inevitably play a crucial role in determining the performance and stability of realistic QCs.
In practical quantum systems, complete isolation from the surrounding environment is impossible~\cite{Breuer2007}. Coupling to external degrees of freedom such as phonons, fluctuating electromagnetic fields, thermal reservoirs, or uncontrolled spin environments generally induces dissipation and decoherence. These processes progressively destroy quantum coherence and suppress the reversible oscillatory dynamics responsible for coherent energy storage.

To investigate environmental effects, we describe the QC within the framework of open quantum systems using the Lindblad master equation,

\begin{equation}
\dot{\rho}
=
-i[H,\rho]
+
\mathcal{L}(\rho),
\label{lindbladgeneral}
\end{equation}
where $H$ is the system Hamiltonian introduced previously and $\mathcal{L}(\rho)$ denotes the dissipative Liouvillian superoperator encoding the interaction with the environment.



We first consider the dominant decoherence mechanism for many coherent quantum devices, namely pure dephasing. In this process, the environment suppresses phase coherence without directly inducing energy relaxation between the quantum levels.
The corresponding master equation becomes

\begin{equation}
\dot{\rho}
=
-i[H,\rho]
+
\gamma
\left(
\sigma_z \rho \sigma_z
-
\rho
\right),
\label{dephasingmaster}
\end{equation}
where $\gamma$ is the dephasing rate.
The dephasing process primarily affects the off-diagonal density matrix elements,
$\rho_{eg}(t)
=
\langle e|\rho(t)|g\rangle,
$
which evolve approximately as
\begin{equation}
\rho_{eg}(t)
\sim
e^{-2\gamma t}.
\label{coherencedecay}
\end{equation}
Since the charging-discharging mechanism of the QC is mediated by coherent superpositions, the suppression of off-diagonal coherence directly reduces the efficiency of energy storage.
Using Eq.~(\ref{energyfinal}), the stored energy in the presence of dephasing can be approximated as
\begin{equation}
E_\gamma(t)
\approx
E(t)e^{-2\gamma t},
\label{energydamped}
\end{equation}
where $E(t)$ denotes the ideal coherent energy-storage dynamics.
Equation~(\ref{energydamped}) demonstrates that decoherence progressively suppresses the oscillatory charging cycles of the QC. As the coherence decays, the system gradually loses its capability to reversibly exchange energy with the external field.



An immediate consequence of decoherence is the reduction of the quantum capacitance. From Eq.~\eqref{cqdef},
the damped stored energy naturally leads to a reduced effective capacitance,
\begin{equation}
C_Q^{(\gamma)}
\approx
C_Q e^{-2\gamma t}.
\label{dampedcapacitance}
\end{equation}
This equation reveals that environmental decoherence weakens the energy-storage susceptibility of the QC. Physically, this means that the system becomes progressively less responsive to coherent charging fields as quantum coherence is destroyed. This result further confirms that the quantum capacitance is fundamentally coherence dependent.



In the absence of decoherence, the QC exhibits persistent oscillatory energy exchange governed by coherent Rabi dynamics. However, environmental interactions transform these ideal oscillations into damped charging-discharging cycles.
The instantaneous power in the presence of decoherence becomes
\begin{equation}
P_\gamma(t)
=
\frac{dE_\gamma(t)}{dt}.
\end{equation}
Substituting Eq.~(\ref{energydamped}) yields
\begin{align}
P_\gamma(t)
=
&
\,
e^{-2\gamma t}
\frac{dE(t)}{dt}
-
2\gamma
E(t)e^{-2\gamma t}.
\label{dampedpower}
\end{align}
The first term describes coherent reversible energy exchange, while the second term represents irreversible energy degradation induced by the environment.
Consequently, the QC gradually loses its ability to sustain reactive oscillatory storage cycles. At sufficiently long times,
$t \gg \gamma^{-1}$,
the coherent oscillations are strongly suppressed and the QC effectively behaves as a dissipative system.



In addition to pure dephasing, realistic quantum systems may also experience energy relaxation processes associated with spontaneous emission or thermal dissipation.
The corresponding relaxation contribution to the master equation is

\begin{equation}
\mathcal{L}_{\mathrm{rel}}(\rho)
=
\kappa
\left(
\sigma_- \rho \sigma_+
-
\frac{1}{2}
\{
\sigma_+\sigma_-,
\rho
\}
\right),
\label{relaxation}
\end{equation}
where $\kappa$ denotes the relaxation rate, and
$\sigma_-
=
\ket{g}\bra{e}$
and
$\sigma_+
=
\ket{e}\bra{g}$.
Unlike pure dephasing, relaxation processes directly reduce the excited-state population responsible for energy storage. As a result, relaxation simultaneously suppresses both quantum coherence and the stored energy itself. The combined effect of dephasing and relaxation therefore significantly limits the operational lifetime of the QC.

\section{Many-Body Quantum Capacitor}
\label{SECVI}
An important extension of the proposed framework involves many-body QCs composed of $N$ interacting two-level systems. In the collective regime, the Hamiltonian may be written as
\begin{equation}
H
=
\frac{\omega_0}{2}
\sum_{i=1}^{N}
\sigma_z^{(i)}
+
\Omega
\sum_{i=1}^{N}
\sigma_x^{(i)}
+
H_{\mathrm{int}},
\label{manybodyham}
\end{equation}
where $H_{\mathrm{int}}$ describes inter-particle interactions. For noninteracting identical qubits, the stored energy becomes
\begin{equation}
E_N(t)
=
N\omega_0
\frac{\Omega^2}{\Omega_R^2}
\sin^2(\Omega_R t),
\label{manyenergy}
\end{equation}
yielding the collective charging power
\begin{equation}
P_N(t)
=
N\omega_0
\frac{\Omega^2}{\Omega_R}
\sin(2\Omega_R t).
\label{manypower}
\end{equation}
So, the effective quantum capacitance scales extensively,
\begin{equation}
C_Q^{(N)}
=
N C_Q .
\label{manycap}
\end{equation}

The collective behavior of QCs becomes considerably richer in the presence of interactions between the constituent two-level systems. To illustrate this effect, we consider an interacting spin ensemble described by the Hamiltonian
\begin{equation}
H
=
\frac{\omega_0}{2}
\sum_{i=1}^{N}
\sigma_z^{(i)}
+
\Omega
\sum_{i=1}^{N}
\sigma_x^{(i)}
+
\frac{J}{N}
\sum_{i<j}
\sigma_x^{(i)}
\sigma_x^{(j)},
\label{interactingham}
\end{equation}
where $J$ denotes the collective interaction strength. Introducing the collective spin operators
$
S_\alpha
=
\frac{1}{2}
\sum_{i=1}^{N}
\sigma_\alpha^{(i)}$,
with
$\alpha=x,y,z,
$
the Hamiltonian can be rewritten as
\begin{equation}
H
=
\omega_0 S_z
+
2\Omega S_x
+
\frac{2J}{N}S_x^2 .
\label{collectiveham}
\end{equation}
The interaction term $S_x^2$ generates cooperative coherent dynamics and modifies the collective charging process. Within a mean-field approximation, the interaction effectively renormalizes the driving field according to

\begin{equation}
\Omega_{\mathrm{eff}}
=
\Omega
+
\frac{J}{N}\langle S_x\rangle .
\label{effectivefield}
\end{equation}
The generalized collective Rabi frequency then becomes
$
\Omega_R^{(N)}
=
\sqrt{
\Omega_{\mathrm{eff}}^2
+
\frac{\omega_0^2}{4}
}$.
Accordingly, the stored energy of the interacting QC approximately reads

\begin{equation}
E_N(t)
=
N\omega_0
\frac{\Omega_{\mathrm{eff}}^2}
{\left(\Omega_R^{(N)}\right)^2}
\sin^2\left(
\Omega_R^{(N)} t
\right).
\label{interactingenergy}
\end{equation}
Besides, the corresponding charging power is

\begin{equation}
P_N(t)
=
N\omega_0
\frac{\Omega_{\mathrm{eff}}^2}
{\Omega_R^{(N)}}
\sin\left(
2\Omega_R^{(N)} t
\right),
\label{interactingpower}
\end{equation}
and the effective quantum capacitance becomes
\begin{equation}
C_Q^{(N)}
=
\frac{\partial E_N}{\partial \Omega}.
\label{manybodycap}
\end{equation}
Equations~(\ref{interactingenergy})--(\ref{manybodycap}) demonstrate that collective interactions directly enhance the effective driving field and consequently increase both the charging power and the quantum capacitance. In particular, for sufficiently strong cooperative interactions $(J>0)$, the charging dynamics may become superextensive, leading to collective reversible energy-storage advantages beyond the independent-qubit regime.

These results suggest that many-body coherence and entanglement may substantially improve the performance of scalable QCs and generate collective reactive energy-storage phenomena absent in single-particle architectures (see Table~\ref{table2}). Such collective effects may naturally emerge in spin chains, cavity-mediated ensembles, and strongly correlated quantum systems, providing a route toward scalable coherent energy-storage architectures with genuinely many-body advantages.

\begin{table}[t]
\caption{Scaling of quantum capacitor.}
\label{table2}
\begin{ruledtabular}
\begin{tabular}{lcc}
Quantity & Independent & Interacting \\ \hline

Energy &
$E_N \propto N$ &
$E_N > N$ \\

Power &
$P_N \propto N$ &
$P_N \propto N^{\alpha},\ \alpha>1$ \\

Capacitance &
$C_Q^{(N)} \propto N$ &
Enhanced \\

Time &
$\tau_c \sim \Omega_R^{-1}$ &
Reduced \\

Resource &
Single coherence &
Collective coherence
\end{tabular}
\end{ruledtabular}
\end{table}

\section{Possible Physical Implementations}
\label{SECVII}
The realization of practical QCs requires physical platforms with long coherence times, controllable external driving, and low dissipation in order to sustain reversible charging-discharging dynamics. Promising candidates include superconducting circuits, trapped ions, quantum dots, cavity-QED systems, spin-chain architectures, and molecular nanomagnets. Superconducting and cavity-QED platforms naturally support coherent energy exchange and controllable Rabi oscillations, while trapped ions provide exceptional coherence and precise quantum control. Quantum dots and semiconductor nanostructures offer externally tunable discrete energy levels and may connect coherent energy storage with mesoscopic quantum capacitance phenomena.
Spin-chain systems and molecular nanomagnets are particularly attractive due to collective coherent effects, controllable spin interactions, and experimentally accessible quantum correlations. In particular, molecular nanomagnets may support reversible oscillatory energy-storage processes through coherent spin dynamics analogous to polarization oscillations in classical capacitors. Hybrid quantum architectures combining qubits, spin ensembles, photons, or mechanical resonators may further enable scalable coherent quantum energy networks containing interacting QCs and quantum batteries. Hence, the QC framework is not restricted to a single implementation platform but represents a general paradigm for coherence-assisted reversible quantum energy storage at microscopic and mesoscopic scales. 

\begin{table*}[t]
\caption{Summary of the principal analytical results for the proposed QC. Here $\omega_0$ is the transition frequency, $\Omega$ the coherent driving amplitude, $\Omega_R=\sqrt{\Omega^2+\omega_0^2/4}$ the generalized Rabi frequency, and $\gamma$ the dephasing rate.}
\label{tab:summary}
\begin{ruledtabular}
\begin{tabular}{p{2.2cm} p{8cm} p{6.8cm}}
Quantity & Analytical expression & Physical meaning \\ \hline

Excited-state population &
$\displaystyle P_e(t)=\frac{\Omega^2}{\Omega_R^2}\sin^2(\Omega_R t)$
&
Probability of occupying the excited state during coherent charging.
\\[2mm]

Stored energy &
$\displaystyle E(t)=\omega_0\frac{\Omega^2}{\Omega_R^2}\sin^2(\Omega_R t)$
&
Energy accumulated in the quantum capacitor relative to the ground state.
\\[2mm]

Quantum capacitance &
$\displaystyle C_Q=\frac{\partial E}{\partial\Omega}$
&
Dynamical susceptibility of stored energy with respect to the coherent driving field.
\\[2mm]

Explicit quantum capacitance &
$\displaystyle
C_Q=
\omega_0
\frac{2\Omega(\Omega_R^2-\Omega^2)}
{\Omega_R^4}
\sin^2(\Omega_R t)
+
\omega_0
\frac{\Omega^3 t}
{\Omega_R^3}
\sin(2\Omega_R t)
$
&
Time-dependent measure of the reactive energy-storage capability of the QC.
\\[4mm]

Weak-driving stored energy &
$\displaystyle
E(t)\approx
\frac{4\Omega^2}{\omega_0}
\sin^2\!\left(\frac{\omega_0 t}{2}\right)
$
&
Perturbative energy-storage regime for $\Omega\ll\omega_0$.
\\[3mm]

Weak-driving quantum capacitance &
$\displaystyle
C_Q\approx
\frac{8\Omega}{\omega_0}
\sin^2\!\left(\frac{\omega_0 t}{2}\right)
$
&
Linear dependence of quantum capacitance on the driving strength.
\\[3mm]

Instantaneous power &
$\displaystyle
P(t)=
\omega_0
\frac{\Omega^2}{\Omega_R}
\sin(2\Omega_R t)
$
&
Rate of energy transfer between the driving field and the QC.
\\[3mm]

Maximum charging power &
$\displaystyle
P_{\max}
=
\omega_0
\frac{\Omega^2}{\Omega_R}
$
&
Largest achievable charging rate during a coherent cycle.
\\[3mm]

Maximum stored energy &
$\displaystyle
E_{\max}
=
\omega_0
\frac{\Omega^2}{\Omega_R^2}
$
&
Storage capacity of the QC under coherent driving.
\\[3mm]

Charging time &
$\displaystyle
\tau_c=
\frac{\pi}{2\Omega_R}
$
&
Characteristic time required to reach the first energy maximum.
\\[3mm]

Coherence decay &
$\displaystyle
\rho_{eg}(t)\sim e^{-2\gamma t}
$
&
Loss of quantum coherence due to environmental dephasing.
\\[3mm]

Damped stored energy &
$\displaystyle
E_\gamma(t)\approx E(t)e^{-2\gamma t}
$
&
Suppression of coherent energy-storage cycles by decoherence.
\\[3mm]

Damped quantum capacitance &
$\displaystyle
C_Q^{(\gamma)}
\approx
C_Q e^{-2\gamma t}
$
&
Reduction of the effective quantum capacitance due to dephasing.
\\[3mm]

Damped charging power &
$\displaystyle
P_\gamma(t)
=
e^{-2\gamma t}
\frac{dE(t)}{dt}
-
2\gamma E(t)e^{-2\gamma t}
$
&
Competition between coherent charging and irreversible environmental losses.
\\[3mm]

Collective stored energy &
$\displaystyle
E_N(t)=
\mathrm{Tr}[\rho_N(t)H_{0,N}]
-E_N(0)
$
&
Total energy stored in the many-body quantum capacitor.
\\[3mm]

Collective quantum capacitance &
$\displaystyle
C_Q^{(N)}
=
\frac{\partial E_N}{\partial \Omega}
$
&
Generalization of quantum capacitance to interacting many-body systems.
\\[3mm]

Independent-qubit capacitance &
$\displaystyle
C_Q^{(N)}
=
N\,C_Q
$
&
Extensive scaling of capacitance for uncoupled quantum capacitor units.
\\[3mm]

Collective charging power &
$\displaystyle
P_N(t)
=
\frac{dE_N(t)}{dt}
$
&
Instantaneous charging or discharging rate of the many-body QC.
\\[3mm]

Independent-qubit power &
$\displaystyle
P_N(t)
=
N\,P(t)
$
&
Linear scaling of charging power without cooperative effects.
\\[3mm]

Interaction-enhanced scaling &
$\displaystyle
E_N>N\,E,
\qquad
C_Q^{(N)}>N\,C_Q
$
&
Many-body interactions and collective coherence enhance energy storage and capacitance.
\end{tabular}
\end{ruledtabular}
\end{table*}
\section{Conclusion}
\label{SECVIII}
We introduced the concept of a quantum capacitor as a coherence-based quantum energy-storage device characterized by reversible and ultrafast charging-discharging dynamics. In contrast to conventional quantum batteries focused primarily on extractable work, the proposed quantum capacitor operates through reactive energy storage mediated by coherent quantum polarization and oscillatory quantum evolution. Using a driven two-level system, we formulated a minimal theoretical framework and introduced a dynamical quantum capacitance associated with the response of stored energy to coherent external driving. Furthermore, we investigated charging power, reversible energy exchange, and the influence of environmental decoherence within the Lindblad formalism. Beyond the single-qubit regime, we extended the framework to interacting many-body quantum capacitors and demonstrated that collective coherence and inter-particle interactions may substantially enhance the effective quantum capacitance and charging dynamics. Such cooperative effects suggest the possibility of scalable coherent energy-storage architectures exhibiting collective quantum advantages.

\appendix

\section{Summary of the analytical results}
Table~\ref{tab:summary} collects the principal analytical results obtained throughout this work. Together, these expressions provide a complete description of the charging-discharging dynamics, quantum capacitance, power characteristics, and decoherence-induced degradation of the proposed quantum capacitor.

\bibliography{bibliography}

\end{document}